\documentclass[10pt,conference]{IEEEtran}
\IEEEoverridecommandlockouts
\usepackage{caption}

\usepackage{times}
\usepackage{cite}

\ifCLASSINFOpdf
  \usepackage[pdftex]{graphicx}
\else
\fi
\usepackage{array}
\usepackage{booktabs}
\usepackage{units}
\usepackage{mhchem}
\begin{document}
\noindent
© 2024 IEEE. Personal use of this material is permitted. Permission from IEEE must be obtained for all other uses, in any current or future media, including reprinting/republishing this material for advertising or promotional purposes, creating new collective works, for resale or redistribution to servers or lists, or reuse of any copyrighted component of this work in other works.

\vspace{1em} 

\title{Development of TiN/AlN-based
superconducting qubit components
}


\markboth{Journal of IEEE Nano}%
{Shell \MakeLowercase{\textit{et al.}}: Development of TiN/AlN-based superconducting\newline qubit components}
\author{
\IEEEauthorblockN{ Benedikt Schoof}
\IEEEauthorblockA{\textit{School of Computation, Information}\\ \textit{and Technology} \\
\textit{Technical University of Munich}\\
Garching, Germany \\
benedikt.schoof@tum.de}
\and
\IEEEauthorblockN{ Moritz Singer}
\IEEEauthorblockA{\textit{School of Computation, Information}\\ \textit{and Technology} \\
\textit{Technical University of Munich}\\
Garching, Germany 
}
\and
\IEEEauthorblockN{ Simon Lang}
\IEEEauthorblockA{\textit{Fraunhofer Institute for Electronic }\\ \textit{Microsystems and Solid State}\\ \textit{ Technologies EMFT} \\
\textit{Fraunhofer EMFT}\\
Munich, Germany \\
}
\and
\IEEEauthorblockN{Harsh Gupta}
\IEEEauthorblockA{\textit{School of Computation, Information}\\ \textit{and Technology} \\
\textit{Technical University of Munich}\\
Garching, Germany \\
}
\and
\IEEEauthorblockN{Daniela Zahn}
\IEEEauthorblockA{\textit{Fraunhofer Institute for Electronic }\\ \textit{Microsystems and Solid State}\\ \textit{ Technologies EMFT} \\
\textit{Fraunhofer EMFT}\\
Munich, Germany \\
}
\and
\IEEEauthorblockN{ Johannes Weber}
\IEEEauthorblockA{\textit{Fraunhofer Institute for Electronic }\\ \textit{Microsystems and Solid State}\\ \textit{ Technologies EMFT} \\
\textit{Fraunhofer EMFT}\\
Munich, Germany \\
}
\and
\IEEEauthorblockN{ Marc Tornow}
\IEEEauthorblockA{\textit{School of Computation, Information} \\
\textit{and Technology} \\
\textit{Technical University of Munich;}\\
\textit{Fraunhofer Institute for Electronic }\\ \textit{Microsystems and Solid State}\\ \textit{ Technologies EMFT} \\
Garching, Germany \\
}
}

\maketitle

\begin{abstract}

This paper presents the fabrication and characterization of superconducting qubit components from titanium nitride (TiN) and aluminum nitride (AlN) layers to create Josephson junctions and superconducting resonators in an all-nitride architecture.
Our methodology comprises a complete process flow for the fabrication of TiN/AlN/TiN junctions, characterized by scanning electron microscopy (SEM), atomic force microscopy (AFM), ellipsometry and DC electrical measurements.
We evaluated the sputtering rates of AlN under varied conditions, the critical temperatures of TiN thin films for different sputtering environments, and the internal quality factors of TiN resonators in the few-GHz regime, fabricated from these films. Overall, this offered insights into the material properties critical to qubit performance.
Measurements of the dependence of the critical current of the TiN / AlN / TiN junctions yielded values ranging from 150 µA to 2 µA, for AlN barrier thicknesses up to ca. 5 nm, respectively.
Our findings demonstrate advances in the fabrication of nitride-based superconducting qubit components, which may find applications in quantum computing technologies based on novel materials.

\end{abstract}


\IEEEpeerreviewmaketitle

\section{Introduction}

\IEEEPARstart{S}{UPERCONDUCTING} qubits stand at the forefront of today's
quantum computing platforms, offering promising pathways to addressing previously unsolvable computational problems.\cite{reviewQuantum} Nowadays, the fabrication of these qubits relies mostly on materials such as niobium and aluminum, each with unique advantages but also significant limitations. The most common configuration for Josephson junctions, the core component of superconducting qubits, involves aluminum electrodes separated by an aluminum oxide barrier.\cite{AluminumJunctionTiN,MoreManhattanJunctions,MoreManhattanJunctions2,aluqubits,niobqubit,tantalqubit}
However, this configuration presents substantial challenges, notably limitations of the coherence time attributed to the aluminum oxide barrier (as host for two-level-systems, TLS, as one of the predominant loss sources) and the instability of aluminum in diluted hydroflouric acid (HF), a common agent, often used to remove native oxides, before mounting the qubit chips into a cryostat.
Titanium nitride (TiN)), a well-established material in CMOS technology, offers a compelling alternative due to its better stability against oxidation and its chemically inert properties at room temperature, contrasting with the more reactive nature of niobium or aluminum.
All-nitride superconducting qubit components, specifically based on AlN as barrier material in Josephson junctions (JJs), have been reported before, mostly using NbN as superconducting electrode material\cite{AlNQubit1,AlNQubit2}.
Recent advancements in superconducting qubit technology have demonstrated the potential of TiN in enhancing qubit coherence times.
For example, a study of TiN films deposited on sapphire substrates reported low dielectric losses at the material interface, achieving qubit lifetimes of up to 300 µs and quality factors approaching 8 million \cite{AluminumJunctionTiN}.
Additionally, the high kinetic inductance of TiN can be utilized to minimize the footprint of qubit components in comparison with low kinetic inductance materials \cite{KinInductance,KinTin1}.
TiN-based JJs have also been reported before, for example, in a nanobridge architecture \cite{nanobridgeTiN}. This paper presents our progress towards all-nitride qubit components, here with a focus on the CMOS compatible materials TiN and AlN, only, with processing steps all transferrable to silicon wafer-scale fabrication, including, e.g., standard lithography, sputter deposition and reactive ion etching. 
By investigating the impact of fabrication processes on the resonator internal quality factors and electrical DC characteristics of TiN/AlN/TiN junctions, we seek to contribute to the development of alternative superconducting qubit platforms.

\section{Materials and Fabrication}
\begin{figure}[htbp]
    \centering
    \includegraphics[width=\linewidth]{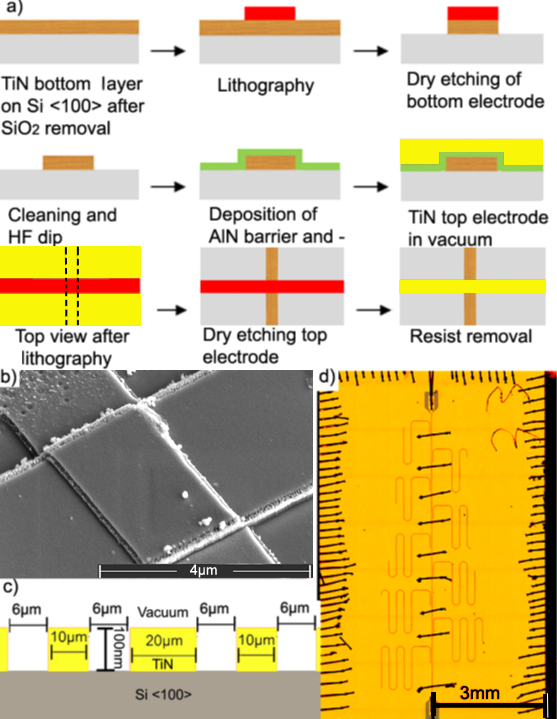}
    \caption{Fabrication and imaging of TiN/AlN/TiN junctions and TiN resonators
    (a) Schematic step-by-step fabrication process of TiN/AlN/TiN junctions.
    (b) SEM image showing the detail of a fabricated junction.
    (c) Schematic, cross-sectional view of the resonator and feedline dimensions, not drawn to scale.
    (d) Laser microscope image of a resonator chip processed at 400°C and 2.5µbar.
    Dark linear features are bonded wire bridges to improve ground-plate potential equalization.}
    \label{fig:figure1}
\end{figure}

The fabrication of TiN resonators and TiN/AlN/TiN Josephson junctions was carried out starting from 4 inch, high-resistance silicon (100), $>$10 kOhm$\cdot$cm, wafers from Topsil, cut into either 4x4 mm$^2$ or 10x6 mm$^2$ chips for junctions and resonators, respectively. 
Initial cleaning involved sequential ultrasonication with acetone and isopropanol (300s each) to remove the protective resist coating, which had been applied before cutting the silicon chips. 
The cleaning is then concluded with a 30-second Buffered Oxide Etch (BOE) 7:1 (NH4F:HF) to strip any native oxides and to hydrogen-terminate the silicon surface. \textit{[caution: hydrofluoric acid (HF(aq)) is very hazardous to health; special care/training is mandatory]}
This ensures a pristine substrate for the subsequent deposition steps. 
All fabrication steps are shown in Fig.\ref{fig:figure1}a.

Bottom Electrode Deposition:
The cleaned substrates were loaded into an Alcatel A 450 RF Magnetron Sputtering Machine, at a base pressure of $\approx$ 1 $\cdot$ $10^{-7}$ torr. 
The titanium nitride (TiN) thin-films for fabricating either the JJ bottom electrodes or the resonator structures were deposited by reactively sputtering from a Ti target in a 10: 1 argon to N2 gas mixture. 
The deposition process was varied between samples at pressures of 4µbar and 2.5µbar, with substrate temperatures set to 673K and room temperature (300K), respectively.
At first, a plasma was initiated at 200 W for a few minutes to stabilize the process before exposing the substrates by opening a shutter, precisely timing the exposure to achieve a consistent film thickness of 120 ± 8 nm, with uncertainty emerging from the exact placement of the substrate in the chamber.
It is known that for higher pressures an increased nitrogen content in the TiN film is obtained, and an increased sputtering temperature leads to a better crystalline order.\cite{TiNComprehensive}

\begin{figure}[htbp]
    \centering
    \includegraphics[width=0.9\linewidth]{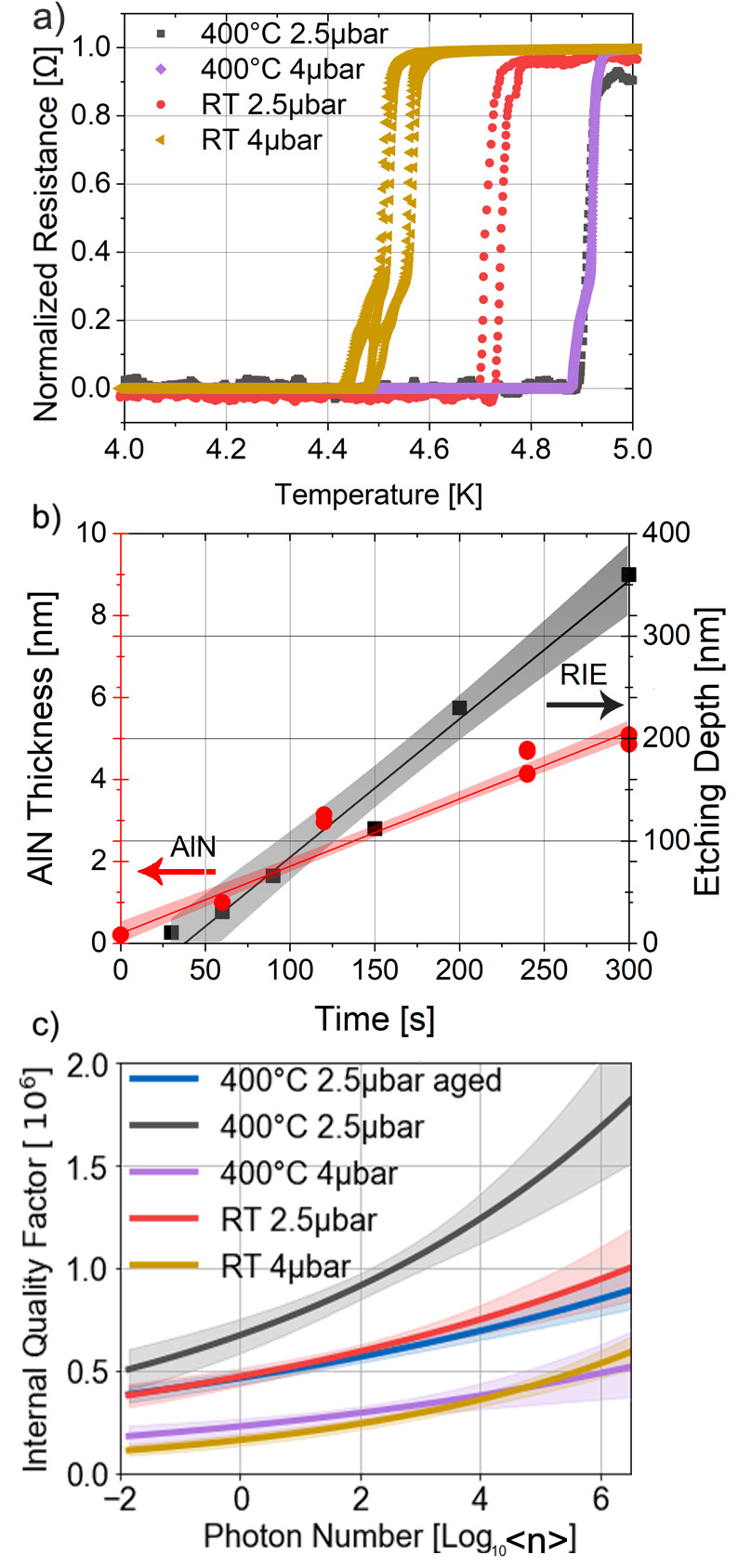}
    \caption{Material properties and performance metrics: 
    (a) Normalized resistance as a function of temperature for 120 nm thick TiN films sputtered on Si $<$ 100 $>$ using different sputtering conditions, measured at constant current 100 µA. The critical temperature increases with increasing sputtering temperature and decreases with increasing pressure.
    (b) AlN sputtering thickness (red) and dry etching depth (black) as a function of time, measured with an AFM. The shaded areas are the 95$\%$ confidence intervals of linear fits (full lines) through the data points.
    (c) Internal quality factor ($Q_i$) at 100 mK as a function of mean photon number $<n>$. The line indicates the mean value for each set of sputtering parameters, averaged over 5 to 8 resonators and the shaded area indicates the 95$\%$ confidence interval, calculated using bootstrapping. The $Q_i$ increases for higher sputtering temperatures and lower pressures.}
    \label{fig:figure2}
\end{figure}

Photolithography and Etching: After deposition, the samples were coated with AZ5214E photoresist, spun at 5000 rpm for 40 seconds and pre-baked at 100 $^\circ$ C for 3.5 minutes.
UV exposure was conducted using a Heidelberg Instruments µMLA (wavelength 365nm at 65mJ/cm²). Development in AZ 400K (1:4 with water) for 40 seconds, followed by a rinse in deionized water, created the desired pattern, which for the resonators was comprising a 'hanger' geometry with 9 resonators capacitively coupled to a feedline.
The patterned TiN films were then etched in a PlasmaPro80 ICP RIE machine, using CF4 gas at 230V bias with a slight targeted overetch to account for chamber variations and to ensure proper structure definition. 
This process was measured to etch 1.25$\pm 0.05\frac{nm}{s}$ after an initial startup phase of $\approx$35 s, as can be seen in Fig. \ref{fig:figure2}b.
After patterning the first TiN layer, the resonator chips were cleaned, BOE dipped, and bonded as illustrated in Fig.\ref{fig:figure1}d, before being loaded into an adiabatic demagnetization cryostat (kiutra, L-Type Rapid).
A cross-sectional view of the resonator can be seen in Fig.1c. The sections, feedline (10µm), coupling plane (20µm) and resonator (10µm) of the structure are separated by 6µm wide gaps. 

Josephson Junction Structuring:
Before depositing the AlN barrier, TiN bottom electrode stripes of width 4µm and length 80µm were patterned from the TiN film analogous to the process described above for the resonators. Subsequently, substrates underwent another round of cleaning and BOE dipping to remove the native oxynitride TiO$_x$N$_y$ that would likely form on TiN. 
AlN was then sputtered at 50W and 11µbar using a 10Ar/15N2 gas mixture.
For the characterization of the sputtering process, a part of the chip was covered with photoresist after the BOE dip.
The AlN on the photoresist was then removed using liftoff.
The step between bare silicon and AlN could then be measured using AFM. By this method, a deposition rate of ca. 1 $\frac{nm}{min}$ was determined, with a variation of about 0.98 $\pm$0.07$\frac{nm}{min}$ (Fig. \ref{fig:figure2}a). 
To account for the possibility of oxidation of the AlN films, thicker AlN $>$ 5 nm was deposited before 20 nm film of TiN was added on top in-situ, without breaking the vacuum. The total thickness was measured in a manner similar to that of the AlN, described above. To measure the thickness of AlN below the TiN, ellipsometry was performed using a Woollam alpha 2.0 ellipsometer. During the fitting process, the total thickness was fixed to the value obtained by AFM. An AlN deposition rate of 0.99 $\pm$0.03$\frac{nm}{min}$ was measured using this method.
Subsequent to the AlN deposition, without breaking vacuum, the TiN top electrode, 60 nm thick, was deposited.
For subsequent patterning of the top electrode structure, analogous photolithographic and etching procedures as described above, were carried out. For the last RIE step, the etching time was chosen to slightly over-etch the top electrode into the bottom electrode. The etch time was chosen to be 100s, which amounts to 81 $\pm$3 nm. This was done to increase the distance between bottom and top electrode at the edge of the junction and have less risks for shorts.

Final Assembly: The junctions, each 4µm x 4µm in area, were then wire-bonded with aluminum wires at  external  bond pads, and the sample chip was loaded  into  the  cryostat  for  testing at low temperatures. Fig.\ref{fig:figure1}b shows a scanning electron microscope (SEM) image of one of these junctions.

\section{Measurement, Results and Discussion}
The performance of the superconducting resonators was evaluated using RF transmission (S21) measurements at 100mK, focusing on the internal quality factor ($Q_i$) as a function of different fabrication parameters. 
Measurements were carried out using a vector network analyzer (VNA), Keysight Technologies model P5002B.
The investigated parameters were the pressure and temperature during the reactive sputtering of TiN films. 
To measure the superconducting transition temperature ($T_c$ for all 4 films, 4-point resistance measurements were performed on planar films while cycling the temperature of the samples from 15K to 300mK and then back up.
For all films, $T_C$ was within a range of about 0.4K, Fig. \ref{fig:figure2}a. 
To assess the uncertainty of $T_C$, several cycles at the maximum speed of 0.3$\frac{K}{min}$ were conducted for the RT 4µbar TiN film.
In our system, during cooling, $T_C$ always appears to be slightly lower and during heating, $T_C$ appears to be higher. 
This is presumably due to a temperature gradient between the measurement device and the sample. 
The uncertainty of the $T_C$ measurement can be estimated from the hysteresis width of these cycles.
The uncertainty from the thermal coupling is estimated to be $\pm$0.03K for the fastest possible ramping speed.
$T_C$ was calculated as the temperature with highest resistance gradient between two data points.
For films sputtered at elevated temperatures, a high $T_c$ of 4.9K was measured.
For the RT films, $T_c$ values of 4.5K and 4.7K were measured.
Despite the films sputtered at high temperatures having similar $T_c$ values, it was found that the performance differed significantly for the $Q_i$ measurements, see Fig.\ref{fig:figure2}c.
Even after aging the 400°C 2.5µbar film for three days in ambient air, it still significantly outperforms its 4µbar counterpart which was measured shortly after the HF dip. 
The measured quality factors correspond to loss tangents in the range $\delta_i$ = 1.5$\cdot$10$^{-6}$ to 6.2$\cdot$10$^{-6}$. Hence, they are slightly higher than those reported in, e.g. \cite{AlTin} ($\delta_i$ = 0.4$\cdot$10$^{-6}$ to 1.1$\cdot$10$^{-6}$). This might be due to differences in TiN film thickness, the used resonator designs and etching recipes.
Tab.~\ref{table:your_label} summarizes the observed relationship between the sputtering conditions and the resulting $T_c$ and the internal quality factor $Q_i$.
The lack of a clear correlation between $T_c$ and $Q_i$ for high $Q_i$ shows that $T_c$ is not an exclusive parameter to describe the overall quality of TiN for use in superconducting qubit components. Furthermore, it was tried to fit a standard TLS model to the data. The fit did not consistently converge. This was most likely due to the reason that the data did not level off for the highest available powers that our measurement setup can provide.\\
The better performance of TiN resonators deposited with presumably lower nitrogen content and at a higher temperature during sputtering, irrespective of the determined critical temperature, suggests a promising avenue for enhancing qubit efficiency by optimizing material composition: specifically, at higher temperatures, a probably larger crystal-size in the polycrystalline TiN film may have resulted in a lower amount of grain boundaries and therefore, less losses at defects and a higher quality factor \cite{TiNNitrogenContentPressure}.

\begin{table}[ht]
\centering
\begin{tabular}{lcccc}
\toprule

Pressure & Temperature & BOE & Q$_i$ [$10^5$]  & $T_c$ [K] \\
\midrule
4µbar       & RT         & \textless{2 h} & 1.6 $\pm$0.3 & 4.5 \\
4µbar & 400$^\circ$C & \textless{2 h} & 2.3 $\pm$0.4 & 4.9 \\
2.5µbar     & RT         & \textless{2 h} & 4.7 $\pm$0.4 & 4.7 \\
2.5µbar & 400$^\circ$C & \textless{2 h} & 6.7 $\pm$0.8 & 4.9 \\
2.5µbar & 400$^\circ$C & 3 Days & 4.6 $\pm$0.3 &4.9 \\
\bottomrule
\addlinespace[0.5em]
\end{tabular}
\caption{Summary of fabrication parameters and their effects on \textit{Q$_i$} and \textit{T$_C$}. The parameters denote the pressure and temperature during the sputtering process, the time between the BOE dip and the loading into the cryostat, the internal quality factor at $<n>$ = 1, and the critical temperature of each sample.}
\label{table:your_label}
\end{table}

\begin{figure}[htbp]
    \centering
    \includegraphics[width=\linewidth]{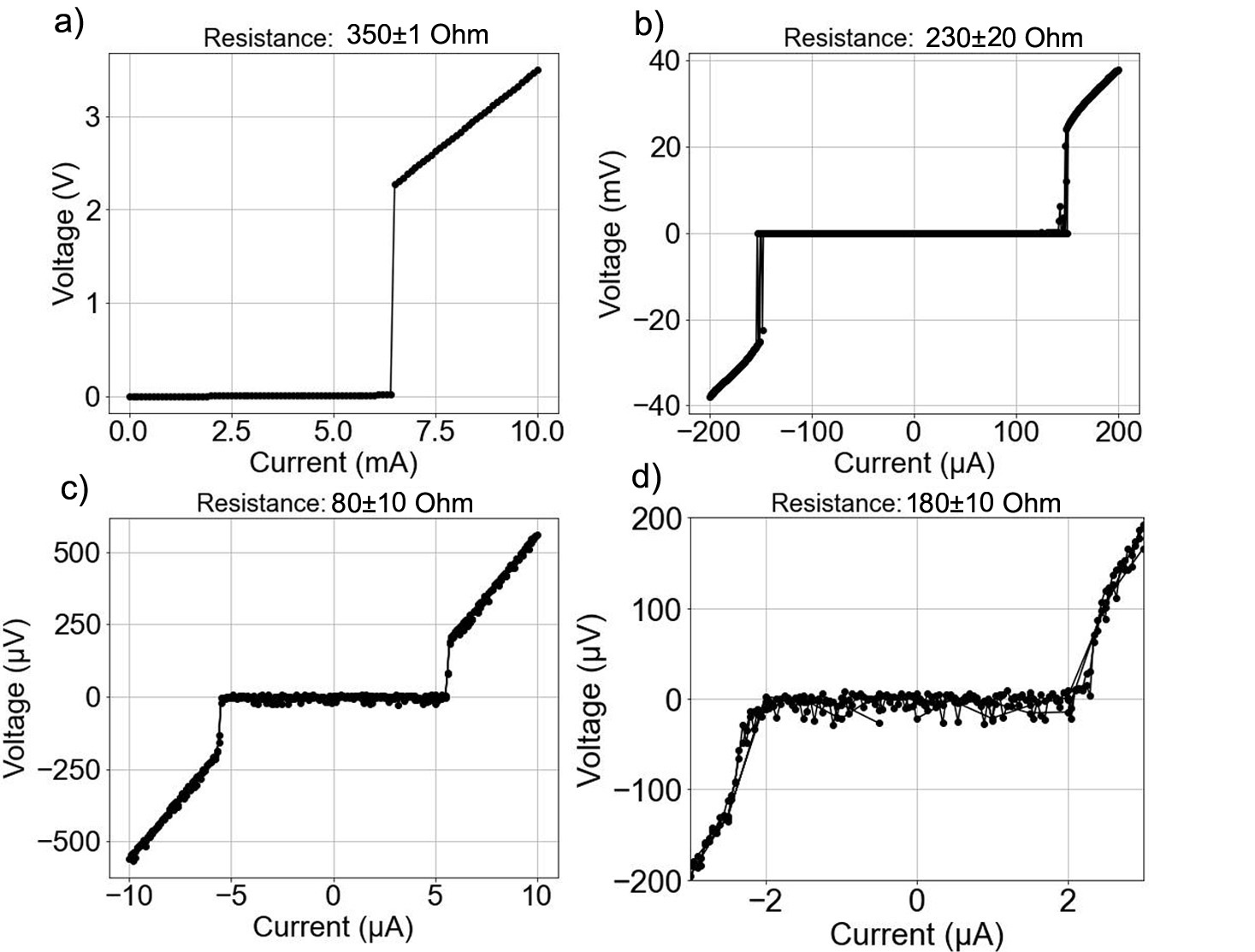}
    \caption{
    Critical current dependency on AlN thickness: Voltage-current characteristics for a junction (a) without barrier (continuous TiN film), indicating the critical current to break the superconducting state, (b) after HF dip of the bottom TiN electrode only, resulting in a native barrier, and (c) with a 2.0 $\pm$ 0.2 nm AlN barrier (d) with a 5.0 $\pm$ 0.5 nm AlN barrier. Resistance values as calculated from the gradient between data points above $I_C$ are denoted above each measurement. All measurements were taken at $T$= 300 mK.}
    \label{fig:figure3}
\end{figure}

For the junctions, we conducted voltage-current ($V-I$) sweeps at a temperature of 0.3K to investigate their DC electrical characteristics and to assess the influence of the AlN barrier thickness on the critical current. 
The temperature was selected to provide a sufficient cooling power margin, which is essential as the sample temperature tended to increase once $I_c$ was reached. 
About half of the tested devices exhibited short circuits, which may be attributed to fabrication artifacts, the relatively large junction area prone to comprise more defects, or to issues arising post-fabrication, such as electrostatic discharges. 
The resulting $V-I$ curves of working junctions showed the expected non-linear behavior characteristic of non-hysteretic (overdamped) Josephson junctions, indicative of the successful fabrication of TiN/AlN-based junctions, see Fig.~\ref{fig:figure3}.

Critical currents extracted from the $V-I$ traces were 150 $\pm$ 10 µA for a native TiOxNy barrier only, 5.7 $\pm$ 0.2 µA  for an approx. 2 nm thick AlN barrier, and 2.5 $\pm$ 0.1 µA for ca. 5 nm AlN, respectively. 
The superconducting energy gap can be calculated using the low-temperature approximated Ambegaokar-Baratoff relation
 \cite{ambegaokar1963tunneling}
\begin{equation}
\Delta(T) = \frac{2eR_N I_c(T)}{\pi},
\label{Amber}
\end{equation}
 with the superconducting energy gap $\Delta$, the resistance in the non-superconducting state $R_N$ and the critical current $I_C$. 
 We here take the slopes of the $V-I$ curves above $I_C$ as values for $R_N$ as first rough approximation, which neglects possible contributions from parallel (intrinsic) shunt resistances \cite{TiNJJintrinsically} that may likely be present in our junctions: $R_N$=230 $\pm$ 20 Ohm for native TiOxNy, $R_N$=80 $\pm$ 10 Ohm for ~2 nm AlN, and $R_N$= 180 $\pm$ 10 Ohm for ~5 nm AlN. 
 For both junctions with known AlN barrier thicknesses (2 nm, 5 nm) $I_C$ and $R_N$ scaled inversely proportional, as expected from Eq. \ref{Amber}, allowing for an estimation of $\Delta$ as calculated from the average of both: 0.29±0.01 meV. This is in the same order of magnitude as the $\Delta$(0) = 0.505 meV ($\hat{=}$0.5 x 8.14 cm$^{-1}$) reported for plasma-enhanced atomic layer deposited TiN \cite{TiNDirectDeltaE}. 
 The value for the sample with native barrier only did not fall into this range, resulting in an unrealistically large value of $\Delta$ (22 $\pm$ 2 meV), which we assign to a not-well-defined (unknown) thickness of the native barrier, which likely also comprised inhomogeneities, so that the Ambegaokar-Baratoff relation may not apply.
 Notably, for 10 nm of AlN, no critical current was measurable at all (data not shown). While in total, our first data on TiN/AlN/TiN junctions indicate a successful fabrication of JJ devices, possible artifacts due to inherent shunt resistances and/or local superconducting pinholes, e.g., at the edges of the crossbar-type µm-scale junctions, cannot be fully excluded and need further investigation.

\section{Conclusion}
In conclusion, our measurements of the superconducting TiN resonators and TiN/AlN/TiN junctions have provided valuable insights into the relevant parameters for the fabrication of all nitride-based superconducting qubits, here focusing on TiN and AlN deposited by sputter processes, only.
By a further careful adjustment of the nitrogen content and crystalline properties in our TiN films together with refining the process for AlN barrier deposition, we anticipate progress toward the fabrication of TiN/AlN-based qubit components in a CMOS-compatible process.
We plan to conduct a comprehensive investigation into the parameters influencing the internal quality factors (\textit{Q$_i$}) of TiN-based resonators. 
This will in particular include systematic measurements of \textit{Q$_i$} as a function of temperature, to study the saturation characteristics of two-level system (TLS) losses.
Upon gaining a deeper understanding of the fabrication and material properties of TiN/AlN/TiN qubit components, our next step will be to develop and characterize functional single qubits using these materials.

\section*{Acknowledgment}
This work was supported by the Munich Quantum Valley project by the Free State of Bavaria, Germany. The authors also express their gratitude to the Walther-Meissner-Institute (Bavarian Academy of Sciences) for providing the script for the resonator layout. Finally, thanks go to Rosemarie Mittermeier and Anika Kwiatkowski of the TUM ZEITlab, for their technical support.

\ifCLASSOPTIONcaptionsoff
  \newpage
\fi



%
\bibliography{IEEEexample}
\bibliographystyle{IEEEtran}

\end{document}